\documentclass[twocolumn,tighten,twocolappendix]{aastex63}

\usepackage{amsmath}
\usepackage{tabularx}

\begin{document}

\title{Can Neutrino Self-interactions Save Sterile Neutrino Dark Matter?}

\shorttitle{Sterile Neutrino DM Constraints}
\shortauthors{An, Gluscevic, Nadler, \& Zhang}

\correspondingauthor{Rui An, Ethan O.~Nadler}
\email{anrui@usc.edu, enadler@carnegiescience.edu}

\author{Rui An}
\affiliation{Department of Physics and Astronomy, University of Southern California, Los Angeles, CA 90089, USA}

\author{Vera Gluscevic}
\affiliation{Department of Physics and Astronomy, University of Southern California, Los Angeles, CA 90089, USA}
\affiliation{TAPIR, Mailcode 350-17, California Institute of Technology, Pasadena, CA 91125, USA}

\author{Ethan O.~Nadler}
\affiliation{Carnegie Observatories, 813 Santa Barbara Street, Pasadena, CA 91101, USA}
\affiliation{Department of Physics and Astronomy, University of Southern California, Los Angeles, CA 90089, USA}

\author{Yue Zhang}
\affiliation{Department of Physics, Carleton University, Ottawa, ON K1S 5B6, Canada}

\begin{abstract}
Sterile neutrinos only interact with the Standard Model through the neutrino sector, and thus represent a simple dark matter (DM) candidate with many potential astrophysical and cosmological signatures. Recently, sterile neutrinos produced through self-interactions of active neutrinos have received attention as a particle candidate that can yield the entire observed DM relic abundance without violating the most stringent constraints from X-ray observations. We examine consistency of this production mechanism with the abundance of small-scale structure in the universe, as captured by the population of ultra-faint dwarf galaxies orbiting the Milky Way, and derive a lower bound on the sterile-neutrino particle mass of 37 keV. Combining these results with previous collider and X-ray limits excludes $100\%$ sterile neutrino DM produced by strong neutrino self-coupling, mediated by a heavy ($\gtrsim 1~\mathrm{GeV}$) scalar; however, data permits sterile-neutrino DM  production via a light mediator.
\end{abstract}

\keywords{\href{http://astrothesaurus.org/uat/96}{Particle astrophysics (96)}; \href{http://astrothesaurus.org/uat/353}{Dark matter (353)}; \href{http://astrothesaurus.org/uat/574}{Galaxy abundances (574)}}

\section{Introduction} 

A neutrino--like particle featuring feeble interactions with the Standard Model has long been an appealing candidate for dark matter (DM)~(\citealt{Gunn:1978gr}). 
However, the three known \textit{active} neutrino states $\nu_a$, produced from the thermal bath in the early universe, are too hot to allow formation of DM structure consistent with observations (e.g.,~\citealt{White:1983fcs}). 
A fourth neutrino $\nu_4$, with mass $m_4$ in the keV range, was thus proposed as a combination of an active component and a new \emph{sterile} neutrino $\nu_s$ (with no Standard-Model interactions), such that
\begin{equation}
|\nu_4\rangle = \cos\theta|\nu_s\rangle + \sin\theta |\nu_a\rangle,
\label{eq:mixing}
\end{equation}
with a small mixing angle $\theta$ (\citealt{Drewes:2016upu,Boyarsky:2018tvu}).

Dodelson and Widrow~(\citealt{1994PhRvL..72...17D}) have shown that $\nu_4$ can be produced non-thermally with the correct relic abundance to constitute all of DM. 
In their proposal, each active neutrino produced through standard interactions could experience active--sterile oscillations, with a small probability of conversion to a nearly-sterile mass eigenstate, i.e., $\nu_4$. 
Oscillations would continue until weak interaction decoupling at a temperature of $\sim$1 MeV, at which point the final relic abundance of $\nu_4$ is set; the relic abundance is thus proportional to the strength of the weak interaction and the mixing angle $\theta$.

The original Dodelson--Widrow (DW) mechanism is, however, in tension with observational constraints. 
In particular, the above active-sterile mixing that leads to sterile-neutrino DM production would also lead to a late-time decay of sterile to active neutrinos, accompanied by X-ray photon production in high-density regions of the universe. 
Non-detection of an X-ray signal in indirect DM searches excludes all mixing angles necessary to produce the full DM relic abundance via the original DW mechanism (\citealt{Boyarsky:2005us,Horiuchi:2013noa,Perez:2016tcq,2020Sci...367.1465D}).

New proposals for sterile--neutrino production have emerged (\citealt{Shi:1998km,Shaposhnikov:2006xi, Asaka:2006ek, Bezrukov:2009th, Nemevsek:2012cd, Roland:2014vba, Dror:2020jzy, Nemevsek:2022anh}), that avoid the radiative decay bounds from X-ray observations.
One such model invokes active neutrino self-interactions (\citealt{2020PhRvL.124h1802D, 2020PhRvD.101k5031K, Kelly:2020aks, Benso:2021hhh}), which have mainly been explored indirectly through the $Z$-boson invisible width~(\citealt{ALEPH:2005ab}), red and flavor-dependent experiments~(\citealt{2018PhRvD..97g5030B, Berryman:2022hds,Esteban:2021tub}). 
A strong neutrino self-interaction is a viable and simple extension of the Standard Model, and could occur via an additional mediator $\phi$ below the weak scale (e.g.~\citealt{Berryman:2022hds}).
This would lead to enhanced out-of-equilibrium production of sterile neutrinos, resulting in sterile-neutrino DM characterized by a phase space distribution that depends on its particle mass, the interaction channel, and the mediator mass. 

\begin{figure}[t!]
\includegraphics[width=0.38\textwidth]{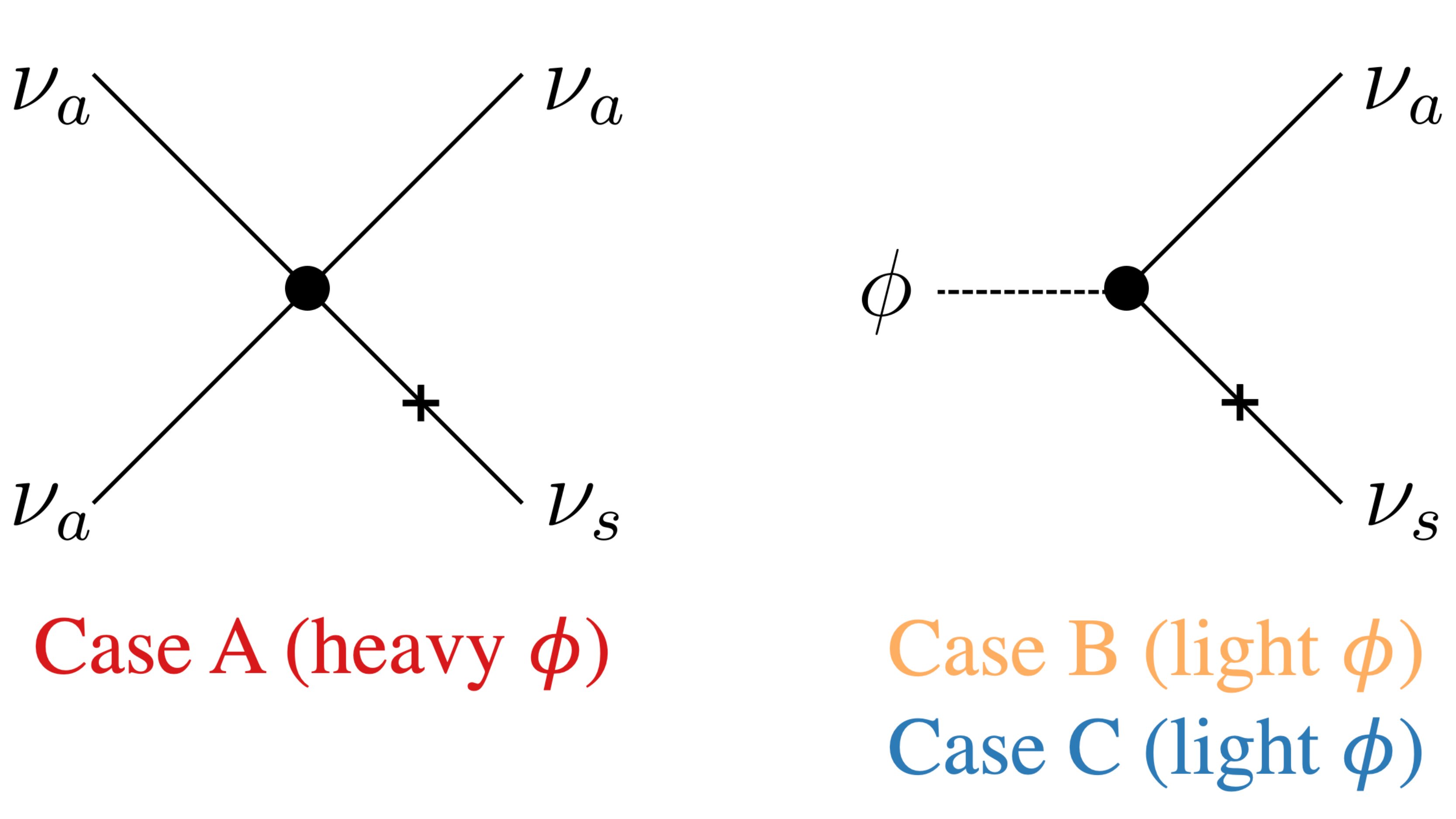}
\centering
\caption{Sterile-neutrino $\nu_s$ production channels, in the presence of active-neutrino self-interactions, in the cases of a heavy (left) or a light (right) scalar mediator. Oscillation of an active into a sterile neutrino is denoted with an ``x". Sterile neutrinos are populated via neutrino self-interactions and mediator decay, respectively, in these two cases.} \label{fig:production}
\end{figure}

For concreteness, we consider the sterile-neutrino production model proposed in \cite{2020PhRvL.124h1802D}, which can populate 100\% of DM in the universe while evading all current constraints.
Within this model, a complex scalar $\phi$ with mass $m_\phi$ mediates new interactions between active neutrinos; here, $G_\chi \equiv \lambda_\phi^2/m_\phi^2$ is the counterpart of the Fermi constant, and $\lambda_\phi$ is the Yukawa coupling between $\phi$ and active neutrinos.
Following \cite{2020PhRvL.124h1802D}, we distinguish three DM production channels (Figure~\ref{fig:production}): production in the presence of neutrino self-interactions through a heavy mediator (Case A); or through an (on-shell) light mediator, with the effective active-sterile mixing angle close to the vacuum mixing angle (Case C) or with a suppressed in-medium mixing (Case B). A number of laboratory and astrophysical constraints on the model were recently explored~(\citealt{Chen:2022kal, Kelly:2021mcd, Kelly:2019wow, 2018PhRvD..97g5030B}).

We confront all three production channels with recent Milky Way (MW) satellite galaxy observations, which place stringent limits on the DM free-streaming scale ~(\citealt{Schneider:2016uqi,Cherry:2017dwu,DES:2020fxi,Zelko:2022tgf}). 
For the first time, we perform probabilistic inference of the free-streaming scale in this class of sterile neutrino DM models using the observed abundance of MW satellite galaxies.
Furthermore, we combine this inference with the upper bound on the active-sterile neutrino mixing angle from X-ray observations and the lower bound on the mixing angle from the invisible Z-boson decay width. It results in a lower bound of 37 keV on the sterile-neutrino DM mass. Our analysis rules out strong neutrino self-interactions (Case A) as a mechanism to produce the entire DM relic abundance.

\section{Model}

We denote sterile-neutrino DM as $|\nu_4\rangle$ and assume that it is a linear combination of active neutrino states $|\nu_a\rangle$ and a new gauge-singlet fermion $|\nu_s\rangle$, such that $\theta$ is the vacuum active--sterile neutrino mixing angle (see Equation~\ref{eq:mixing}).
We further assume that active neutrinos feature self-interactions of the form $\lambda_{\phi} \nu_a\nu_a \phi$.

The key quantity that controls observables in this model is the primordial phase space distribution of sterile-neutrinos $f_{\nu_s}$, which satisfies the following Boltzmann equation (\citealt{2020PhRvL.124h1802D})
\begin{equation}\label{eq:Boltzmann}
\begin{split}
\frac{\mathrm{d} f_{\nu_s}}{\mathrm{d}\widetilde{z}}  &= \frac{\Gamma}{4H\widetilde z} \sin^22\theta_\mathrm{eff} f_{\nu_a},\\
\sin^22\theta_\mathrm{eff} &\simeq \frac{\Delta^2 \sin^22\theta}{\Delta^2 \sin^22\theta + \Gamma^2/4 + (\Delta \cos2\theta - V_T)^2 }.
\end{split}
\end{equation}
We emphasize that $f_{\nu_s}$ needs \textit{not} be a thermal distribution; however, for the parameter space we explore, DM is mostly produced before neutrino decoupling, ensuring that the active neutrinos and photons share the same temperature, $T_\gamma=T_{\nu_a}$.
In the above equations, $\widetilde z\equiv 1\mathrm{ MeV} / T_\gamma\propto 1/z$, is inversely proportional to cosmological redshift $z$,
$H$ is the Hubble expansion parameter, $f_{\nu_a}$ is the Fermi-Dirac distribution of active neutrinos, $\Gamma$ is the interaction rate of active neutrinos in the early universe that receives contributions from the standard weak interactions and the $\phi$-mediated self-interactions, and $\theta_{\mathrm{eff}}$ is the reduced (effective) mixing angle within the plasma that features neutrino self-interactions. Note that the vacuum oscillation frequency is given by $\Delta = m_4^2/(2E)$, where $E$ is the neutrino energy; $V_T$ is the thermal potential experienced by active neutrinos, and $m_4$ is the DM mass.

At high temperatures (early times), when $\Gamma, V_T \gg \Delta$, DM is \textit{not} produced due to collisional damping of the effective mixing angle $\theta_{\mathrm{eff}}$; in other words, neutrino interactions, which only involve the active species, ``reset the clock" too frequently to allow for an active neutrino to convert into a sterile neutrino between consecutive scatterings. Similarly, at low temperatures (late times), when $m_\phi\gg T_\gamma$, the production also ceases because all neutrino interactions drop out of equilibrium, and DM abundances are unchanged. 
The DM phase space distribution $f_{\nu_s}$ is thus calculated by integrating the above equations over an intermediate range of redshift, where DM production occurs most efficiently.

\begin{figure*}
\includegraphics[width=0.32\textwidth]{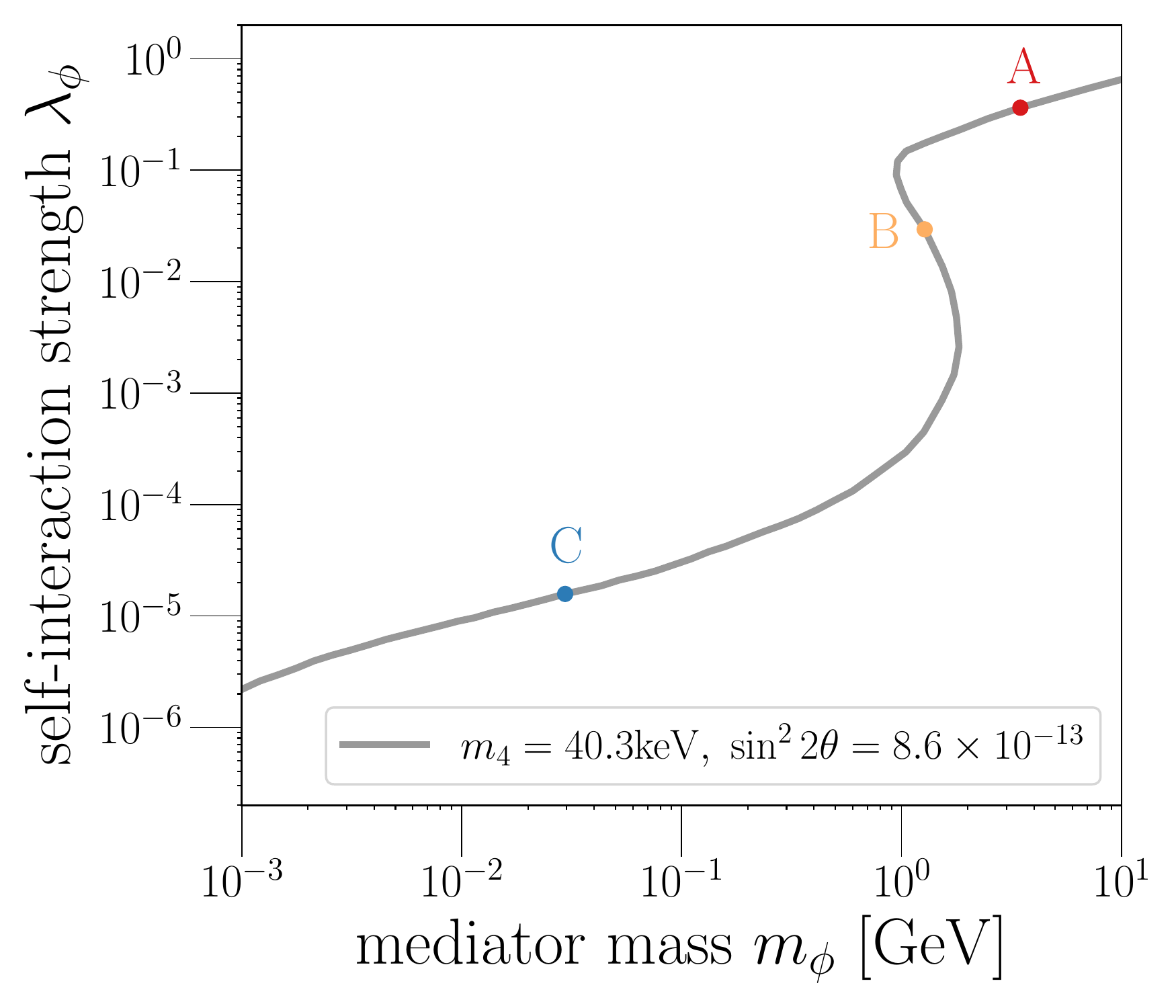}
\includegraphics[width=0.32\textwidth]{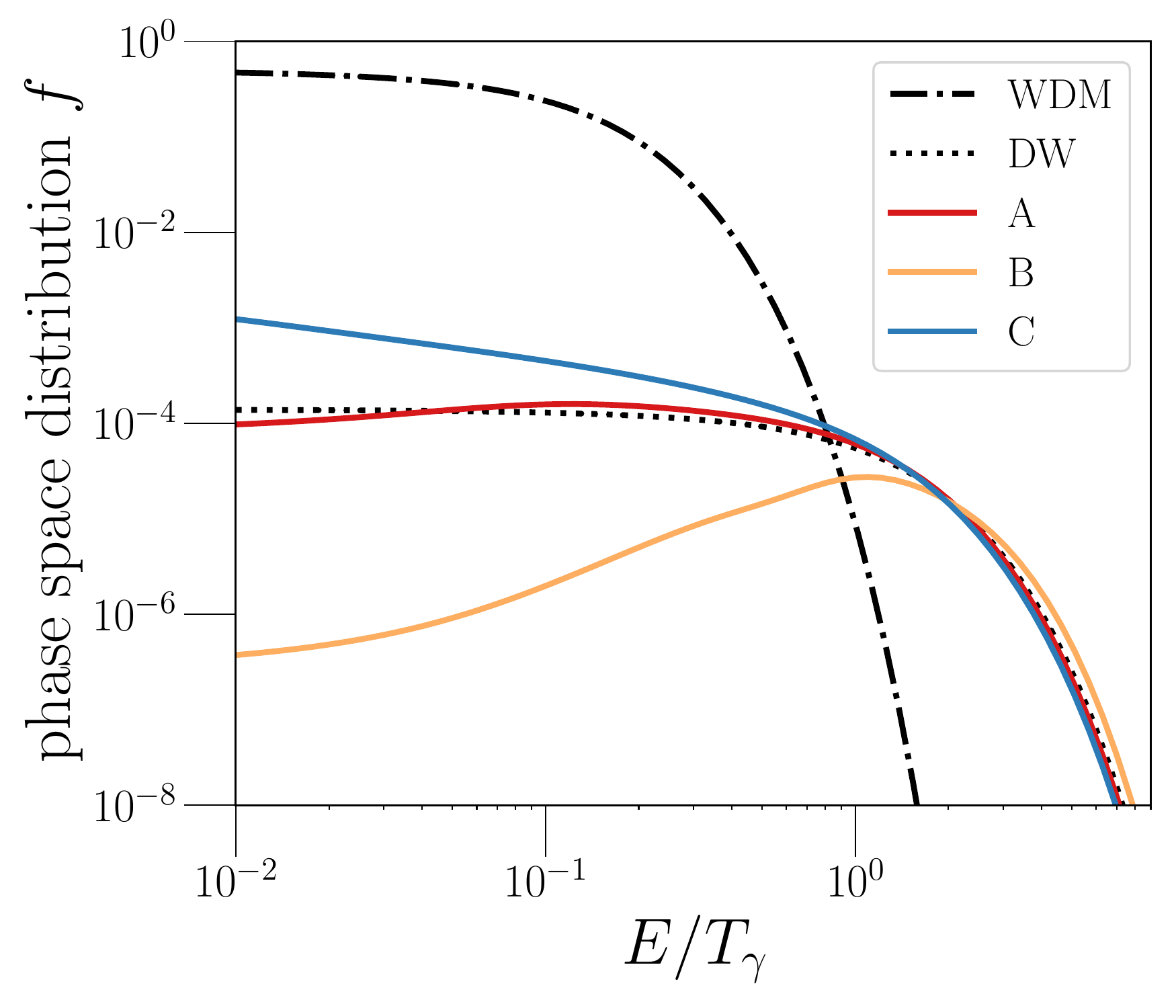}
\includegraphics[width=0.32\textwidth]{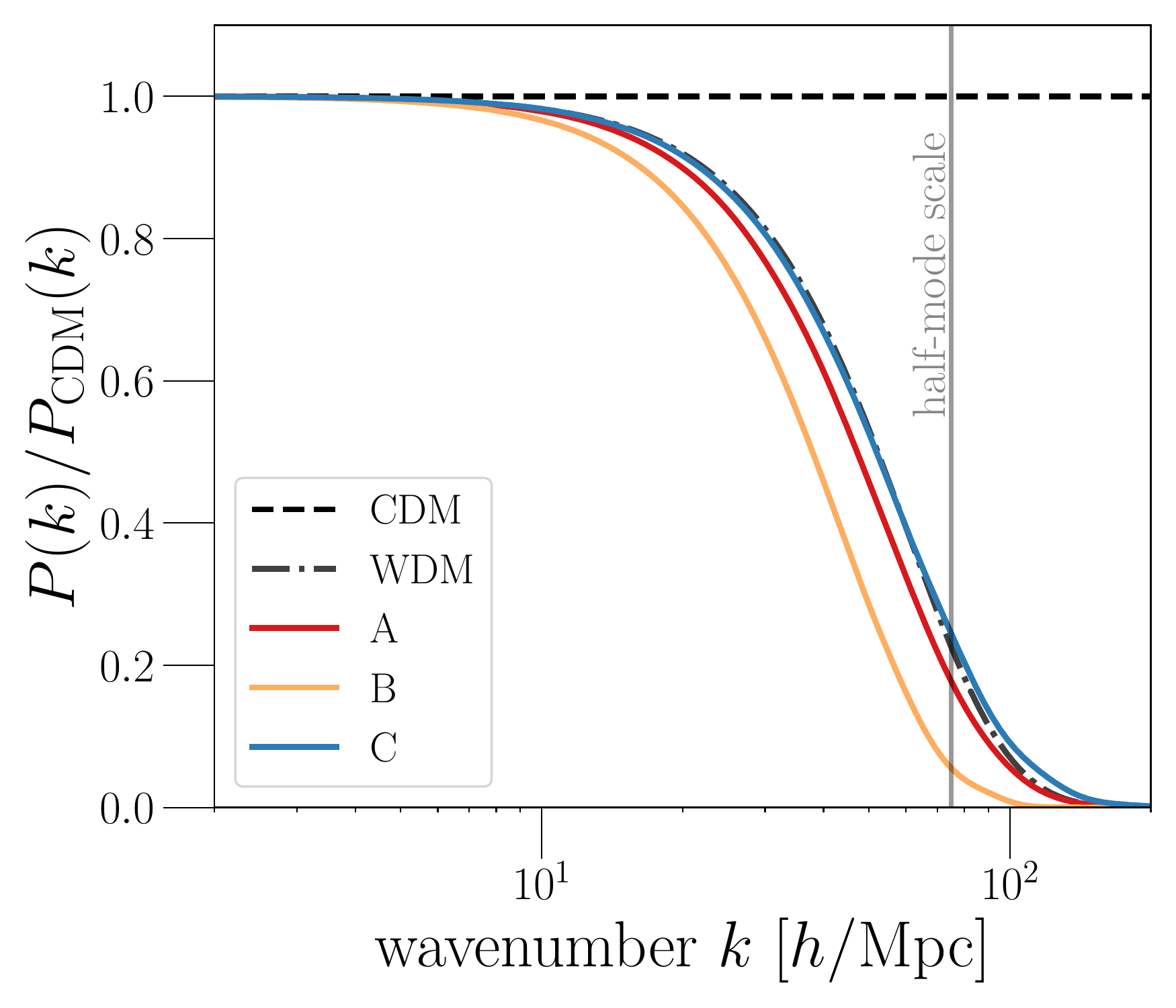}
\caption{\textit{Left panel}: Combinations of neutrino self-interaction strength $\lambda_{\phi}$ and mediator mass $m_{\phi}$ that produce sterile-neutrino DM at the correct relic abundance, for a DM mass of $m_4 = 40.3~\mathrm{keV}$ and vacuum mixing angle of $\sin^2 2\theta = 8.6\times 10^{-13}$. Points in the three DM production regimes are labeled with A, B, and C, corresponding to the Cases described in the text. \textit{Middle panel}: The corresponding DM phase space distributions. For comparison, we also show the original DW model with no neutrino interactions as a dotted line, and the thermal-relic WDM model as a dot-dashed line (evaluated at the $95\%$ confidence $6.5~\mathrm{keV}$ WDM mass bound from \cite{DES:2020fxi}. \textit{Right panel}: The corresponding suppression of DM clustering is illustrated as the ratio of the linear matter power spectrum in sterile-neutrino DM models to that of CDM. The vertical line indicates the half-mode scale $k_{\mathrm{hm}}$ of the ruled-out WDM model.}
\label{fig:psd_tk}
\end{figure*}

In the left panel of Figure~\ref{fig:psd_tk}, the $\mathcal{S}$-shaped gray curve illustrates the relationship between the strength of the neutrino self-interactions and the mediator mass, for models that yield the entire DM relic abundance, at a fixed DM particle mass of $m_4 = 40.3~\mathrm{keV}$ and a fixed mixing angle of $\sin^2 2\theta = 8.6\times 10^{-13}$.
The middle panel of Figure~\ref{fig:psd_tk} shows the DM phase space distributions corresponding to the three DM production cases: in Case A (i.e., points with $\lambda_{\phi}\approx 1$ and $m_{\phi}\gtrsim 1~\mathrm{GeV}$, to the right of the right-most ``knee'' in the $\mathcal{S}$-shaped curve), DM is produced through scattering of active neutrinos; in Cases B and C (i.e., points with $\lambda_{\phi}\ll 1$ and $m_{\phi} \lesssim 1~\mathrm{GeV}$ that lie along the roughly vertical segment and along the left-most portion of the $\mathcal{S}$-shaped curve, respectively), DM is produced through the decay of on-shell $\phi$ (\citealt{2020PhRvL.124h1802D}).

Neutrino interactions affect the phase-space distribution differently in the three cases, which generally results in different free-streaming lengths. 
The specifics of the free-streaming in individual models lead to notable differences in DM clustering on small physical scales; the resulting suppression of the linear matter power spectrum, relative to cold DM (CDM) cosmology is illustrated in the right panel of Figure~\ref{fig:psd_tk}. 
Below, we use this suppression of small-scale power to assess the viability of self-interacting sterile-neutrino production regimes.

\section{Analysis} 

To confront the sterile neutrino model with recent small-scale structure constraints, we use the latest public version of the Boltzmann
solver \texttt{CLASS}~(\citealt{2011arXiv1104.2932L, class2011}) to generate the linear matter power spectrum $P(k)$ consistent with a given sterile-neutrino cosmology (where $k$ is the cosmological wavenumber), and compare it to standard CDM cosmology.
For standard cosmological parameters, we use the values consistent with \cite{DES:2020fxi} (see also \citealt{2021ApJ...907L..46M}): Hubble constant $h=0.6932$, present-day DM density parameter $\Omega_\mathrm{dm}h^2 = 0.1153$, present-day baryon density parameter $\Omega_\mathrm{b}h^2 = 0.02223$, amplitude of the scalar perturbations $A_s = 2.464\times 10^{-9}$, scalar spectral index $n_s=0.9608$, and optical depth to reionization $\tau_\mathrm{reio}=0.081$.
To account for the effects of sterile-neutrino free-streaming on $P(k)$, we provide the correct initial phase-space distribution (when DM is still relativistic) as an input to \texttt{CLASS}, for each model parameter combination $\{m_4,\theta,\lambda_{\phi},m_{\phi}\}$. 
We sample over pairs of DM mass and mixing angle $\{m_4,\theta\}$, and choose the interaction strength and the mediator mass $\{\lambda_{\phi},m_{\phi}\}$ that yield a DM relic density consistent with \textit{Planck}.

The key output from \texttt{CLASS} used in our analysis is the ratio of the linear power spectrum for sterile-neutrino DM and CDM cosmologies, $P(k)/P_{\mathrm{CDM}}(k)$. 
This ratio is set at early times and evolved non-linearly as in CDM cosmology; thus, the effects of sterile-neutrino DM on structure formation amount to a change in the initial conditions, shown in the right panel of Figure~\ref{fig:psd_tk}. From this figure, we can see that all three production cases resemble CDM at large scales, $k\lesssim 10~h/\mathrm{Mpc}$, and therefore successfully reproduce large-scale structure observations.
On small scales, power is strongly suppressed due to particle free streaming, leading to fewer low-mass DM halos compared to CDM.

We use the population of satellite galaxies orbiting the MW, as observed by the Dark Energy Survey (DES; \citealt{2018ApJS..239...18A}) and Pan-STARRS1 (PS1; \citealt{2016arXiv161205560C}); these galaxies inhabit DM halos with masses down to $\sim 10^8~M_{\mathrm{\odot}}$~(\citealt{Jethwa:2016gra,DES:2019ltu}), which grew from modes $k\gtrsim 30~h/\mathrm{Mpc}$ (\citealt{Nadler:2019zrb}).
Converting these observations into constraints on any new physics that affects the linear matter power spectrum requires forward-modeling of the satellite population starting from $P(k)$. 
Previous studies performed high-resolution zoom-in simulations of MW-like halos using $P(k)$ for thermal-relic WDM as an initial condition, thus predicting subhalo and galaxy populations in cosmologies with power suppression (\citealt{Schneider:2011yu,Lovell:2013ola,Bose:2016irl,DES:2020fxi,Lovell:2013ola}); these results were used to bound the thermal-relic WDM particle mass (\citealt{DES:2019vzn,DES:2019ltu,DES:2020fxi}). 

To obtain sterile-neutrino bounds, we map linear power spectra for our model, illustrated in the right panel of Figure~\ref{fig:psd_tk}, to the thermal-relic WDM constraint from MW satellite galaxies of~$m_{\mathrm{WDM}}>6.5~\mathrm{keV}$ at $95\%$ confidence~(\citealt{DES:2020fxi}). 
We parameterize the $P(k)$ cutoff for each evaluation of our model using the half-mode scale $k_{\rm{hm}}$, defined as the wavenumber where $P(k)$ drops to 25\% of its CDM value, and match to a thermal-relic WDM model with the same $k_{\rm{hm}}$. 
This procedure is accurate because the shape of $P(k)$ in our model is extremely similar to that in WDM (Figure~\ref{fig:psd_tk}, right panel).
We then translate the WDM posterior distribution from the MW satellite analysis in \cite{DES:2020fxi} into a marginalized posterior probability distribution on $m_4$, $\sin^2 2\theta$, and $\lambda_\phi$; the remaining degree of freedom, $m_{\phi}$, is fixed by the DM relic abundance constraint.
The results we leverage here conservatively marginalize over uncertainties in the MW host halo mass and the galaxy--halo connection, and carefully account for the recent merger history of the MW and observational biases of the surveys we consider (see the Appendix and \cite{DES:2020fxi} for details). 

\begin{figure}[t]
\includegraphics[width=0.46\textwidth]{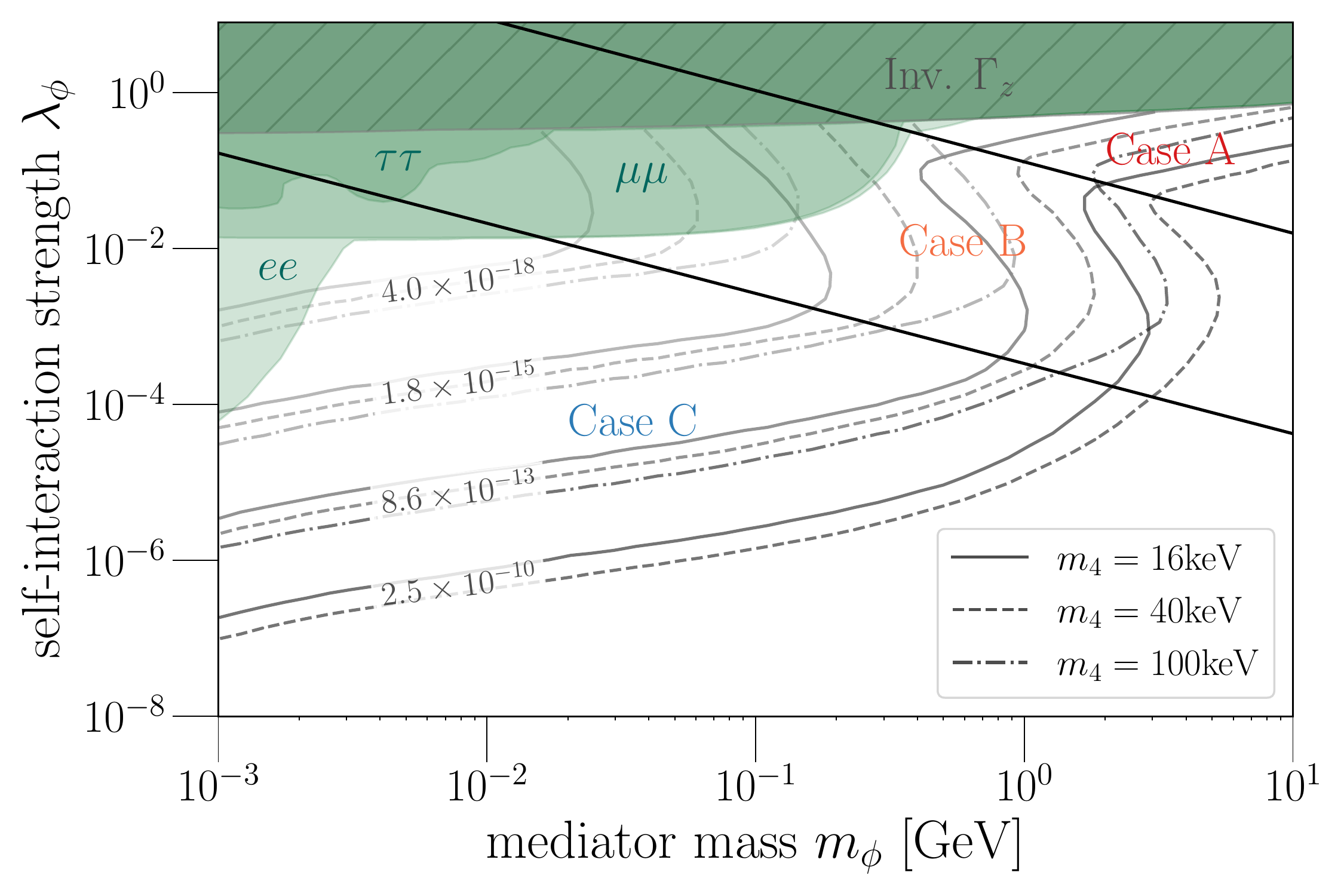}
\centering
\caption{
Neutrino self-interaction parameters that can populate sterile neutrinos with the entire DM relic abundance, for several combinations of $m_4$ and $\sin^2 2\theta$ (gray curves). 
The dark green hatched region shows the constraint from invisible $Z$ width; 
light green shaded regions show experimental flavor-dependent constraints~(\citealt{2018PhRvD..97g5030B, Berryman:2022hds,Esteban:2021tub}).
Diagonal black lines divide this parameter space into production regimes corresponding to Cases A, B, and C.}\label{fig:mphi_lphi}
\end{figure}

\begin{figure*}
\includegraphics[width=0.32\textwidth]{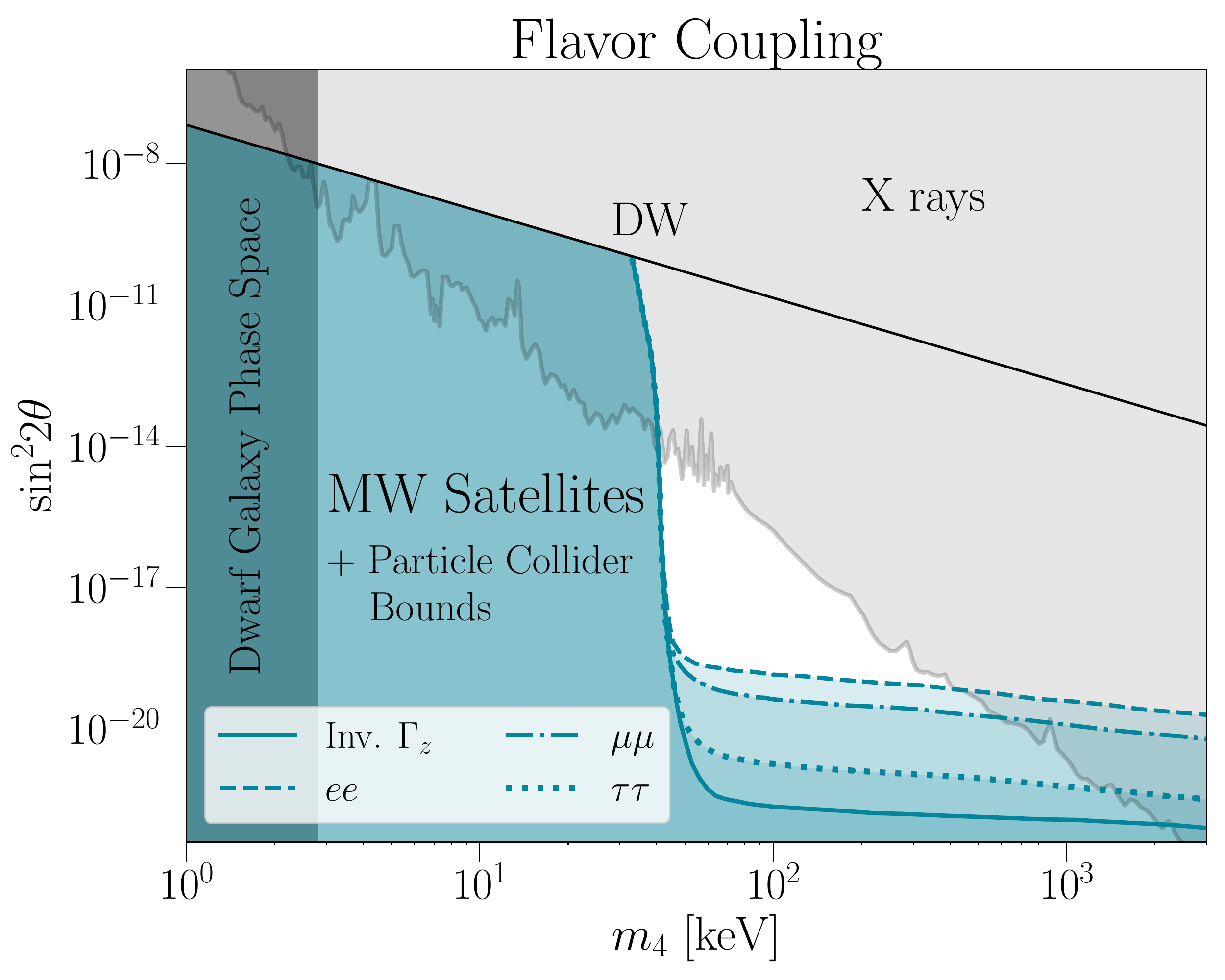}
\includegraphics[width=0.32\textwidth]{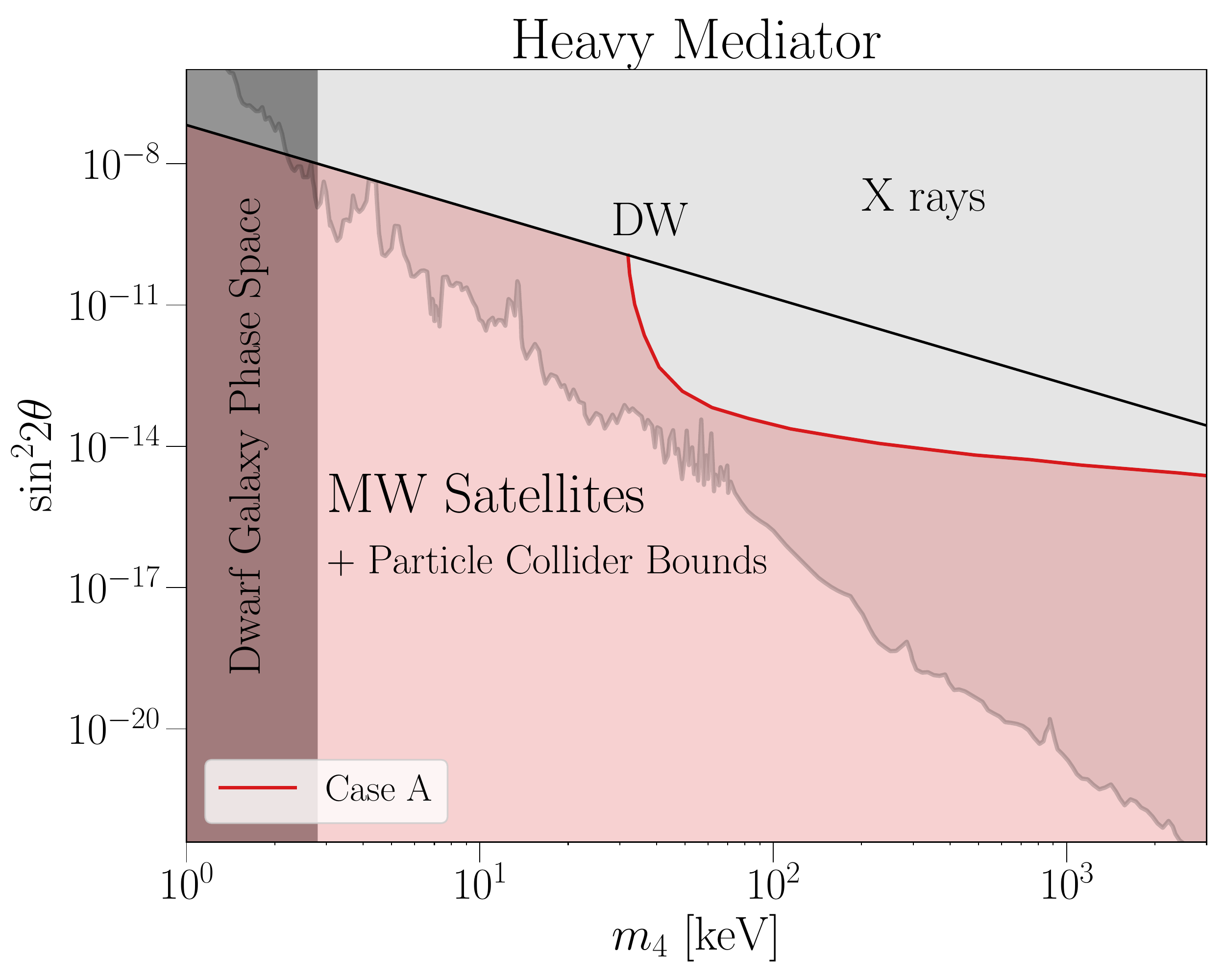}
\includegraphics[width=0.32\textwidth]{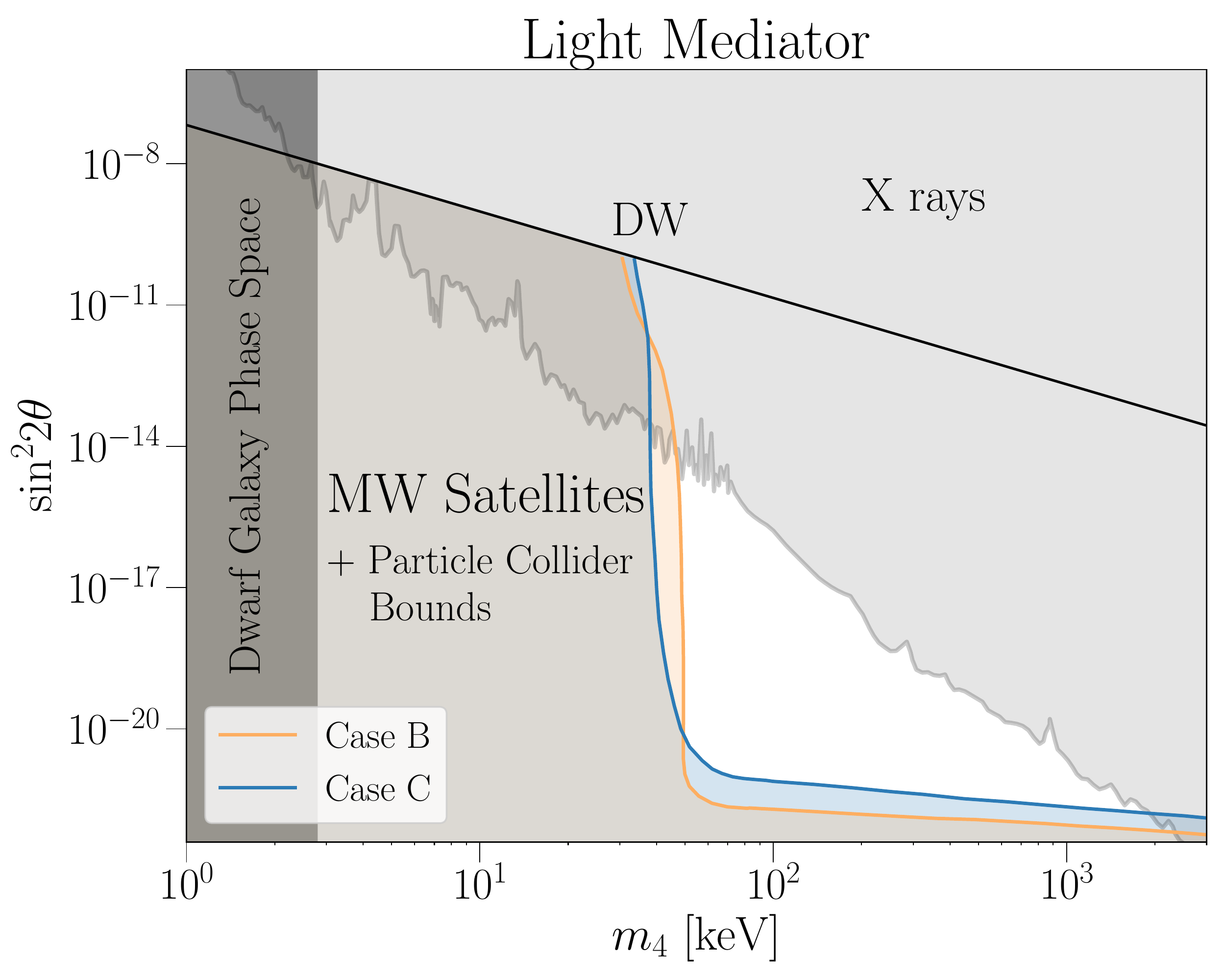}
\caption{
\emph{Left panel}: 95\% confidence exclusion for sterile neutrino DM from our analysis of the MW satellite galaxy population combined with particle collider bounds (colored shaded regions) is compared with X-ray annihilation bounds (light gray; \citealt{2012JCAP...03..018W,Horiuchi:2013noa,2014ApJ...789...13B,2014PhRvL.113y1301B,Perez:2016tcq,2019PhRvD..99h3005N,2020Sci...367.1465D}) and the bounds from dwarf galaxy phase-space densities (dark gray; \citealt{Alvey:2020xsk}). Galaxy--halo connection and neutrino self-interaction parameters ($m_{\phi}$ and $\lambda_{\phi}$) are marginalized over. The blue--green region with solid edges corresponds to the case where we only consider the limits from invisible Z-decays on $\lambda_{\phi}$. The dashed, dash-dot or dotted lines correspond to bounds on specific mediator flavor-coupling scenarios. DW production mechanism corresponds to the solid black line. \textit{Middle and right panels}: Our sterile neutrino DM constraints divided according to the production scenario. Case A (red) is the heavy-mediator scenario ruled out by our analysis. Case B (yellow) and Case C (blue) correspond to light-mediator scenarios.}\label{fig:constraints}
\end{figure*}

We combine the resulting lower limit on sterile-neutrino DM mass with existing constraints on the neutrino self-interaction strength from experiments involving charged $K, \pi$ meson decays, neutrinoless double-beta decay, the $Z$-boson invisible width~(\citealt{2018PhRvD..97g5030B, Berryman:2022hds}), and the astrophysical neutrino observations from the IceCube experiment~(\citealt{Esteban:2021tub}). 
To avoid impacting Big Bang Nucleosynthesis, we only focus on mediator masses above $1\mathrm{MeV}$ scale~(\citealt{2020EPJC...80..294E, 2018PhR...754....1P}).
The regions in the $\lambda_\phi$--$m_\phi$ plane excluded by these experiments are shaded green in Figure~\ref{fig:mphi_lphi}; note that some of the constraints are neutrino-flavor dependent, as labeled.
The $\mathcal{S}$-shaped gray curves in Figure~\ref{fig:mphi_lphi} illustrate the loci of the correct DM relic abundance, for several choices of $m_4$ and $\sin^2 2\theta$. 
We label the three distinctive production regimes~(\citealt{2020PhRvL.124h1802D}), which are roughly divided by the diagonal black lines. These correspond to a heavy and strongly-coupled mediator (Case A), light and feebly-coupled mediator (Case C), and an intermediate regime (Case B), from top to bottom.

\section{Results}

One of our key results is a lower bound on the mass of sterile-neutrino DM, derived mainly from DES and PS1 MW satellite observations, in the presence of neutrino self-interactions. By marginalizing over all other model parameters (see the Appendix for more details), we obtain a $95\%$ confidence limit of
\begin{equation}
m_4 > 37~\mathrm{keV},
\end{equation}
regardless of the production mechanism.

The left panel of Figure~\ref{fig:constraints} shows experimental and observational constraints in the $\sin^22\theta$--$m_4$ plane, where all other parameters are marginalized over; the blue--green shaded regions are ruled out by our combined analysis, while the gray shaded regions are excluded by X-ray observations~(\citealt{2012JCAP...03..018W,2014PhRvD..89b5017H,2014ApJ...789...13B,2014PhRvL.113y1301B,2017PhRvD..95l3002P,2019PhRvD..99h3005N,2020Sci...367.1465D}; also see \citealt{Roach190809037,Foster210202207,Sicilian220812271}) and DM phase-space density measurements in dwarf galaxies~(\citealt{Alvey:2020xsk}). Neutrino self-interactions allow the DM relic abundance to be saturated at lower mixing angles than the original DW production mechanism.

The vertical part of each parameter boundary (i.e., the lower bound on $m_4$) is driven by the observed MW satellite abundance, whereas the horizontal part (i.e., the lower bound on $\theta$) is dominated by the experimental bounds from Figure~\ref{fig:mphi_lphi}. 
In the left panel, we allow all three sterile-neutrino production channels by marginalizing the posterior over all mediator masses and coupling strengths; we also show the impact of the three flavor-specific experimental bounds on neutrino coupling with $\phi$, for electron ($\nu_e$), muon ($\nu_\mu$), and tau ($\nu_\tau$) neutrinos as the dashed, dot-dashed, and dotted boundaries, respectively.
We find that the coupling to $\nu_e$ most strongly constrains the mixing angle for the parameter space that allows the standard neutrino to constitute all of DM. 

In the middle and right panels of Figure~\ref{fig:constraints}, we show the combined observational and experimental constraints in the same parameter space, but this time showing the individual production channels, A, B, and C, separately, as labeled.
This allows us to assess the validity of individual production channels.
In these two panels, we only incorporate the experimental bounds on the invisible $Z$ width, which is flavor-independent and thus conservative.
We find that the parameter space for Case A---sterile-neutrino DM production in the presence of a heavy mediator with $m_{\phi}\gtrsim\,$GeV and a large coupling, $\lambda_{\phi} \gtrsim \mathcal{O}(0.1)$---is entirely ruled out.
In contrast, Cases B and C allow for smaller $\lambda_\phi$, where the corresponding $Z$-decay constraints are much weaker. A small portion of this parameter space (i.e., the white triangle-shaped region in the right panel of Figure~\ref{fig:constraints}) remains unconstrained for DM masses from 37 keV up to several MeV, depending on the other interaction parameters.

\section{Discussion and Conclusions.} 

Our analysis conservatively marginalizes over astrophysical uncertainties that govern the connection between MW satellite galaxies and DM halos, observational uncertainties related to the known population of MW satellite galaxies, and the properties of the MW system, and accounts for the specific phase-space distribution of sterile-neutrino DM produced by active neutrino self-interactions. The fact that we consider active neutrino self-interactions is crucial because the physics we probe is also directly tested by particle collider experiments, unlike many other sterile neutrino DM scenarios (\citealt{2017PhRvL.119y1305H,Bringmann:2022aim,2019PhRvD.100b3533J}). Our results rule out sterile-neutrino DM produced through neutrino interactions mediated by a heavy scalar particle, at $\gg 95\%$ confidence, and constrains sterile-neutrino DM masses between 37 keV to several MeV, regardless of the other production parameters. These are the first bounds derived from a combined analysis of structure formation and collider data and represent the most stringent sterile-neutrino DM constraints to date.

Importantly, unlike previous small-scale structure constraints on other sterile neutrino production mechanisms (e.g., \citealt{Zelko:2022tgf, Schneider:2016uqi}), we do not simply construct a one-dimensional mapping to thermal-relic WDM limits; rather, we cast our self-interacting sterile neutrino DM model into the full posterior distribution from the MW satellite analysis in \cite{DES:2020fxi}, which allows us to carefully examine degeneracies among sterile neutrino DM parameters and with astrophysical parameters governing the galaxy--halo connection. Furthermore, we extract information for specific physical scenarios from the posterior distribution and thus differentiate the mediator flavor-coupling and production regimes shown in the panels of Figure~\ref{fig:constraints}.

Our study informs the interpretation of astrophysical and particle accelerator searches for neutrino self-interactions and sterile neutrino DM. For example, if a heavy, neutrinophilic scalar is discovered in future accelerator neutrino experiments, following recent proposals~(e.g., \citealt{Kelly:2019wow,Kelly:2021mcd}), our results imply that such a scalar \emph{cannot} produce sterile neutrinos that saturate the DM relic abundance. Meanwhile, if future astrophysical searches discover an indirect detection signal consistent with sterile neutrino DM that possesses a small mixing angle, then light neutrino self-interaction mediators should be targeted by terrestrial experiments as a potential ``smoking gun'' of the DM production mechanism.

\begin{acknowledgments}

We thank Alex Drlica-Wagner for comments on the manuscript. VG and RA acknowledge the support from NASA through the Astrophysics Theory Program, Award Number 21-ATP21-0135. VG also acknowledges support from the National Science Foundation under Grant No. PHY-2013951. YZ is supported by the Arthur B. McDonald Canadian Astroparticle Physics Research Institute.

\end{acknowledgments}

\bibliographystyle{yahapj}
\bibliography{ref}

\clearpage

\begin{figure*}[t!]
\includegraphics[width=0.98\textwidth]{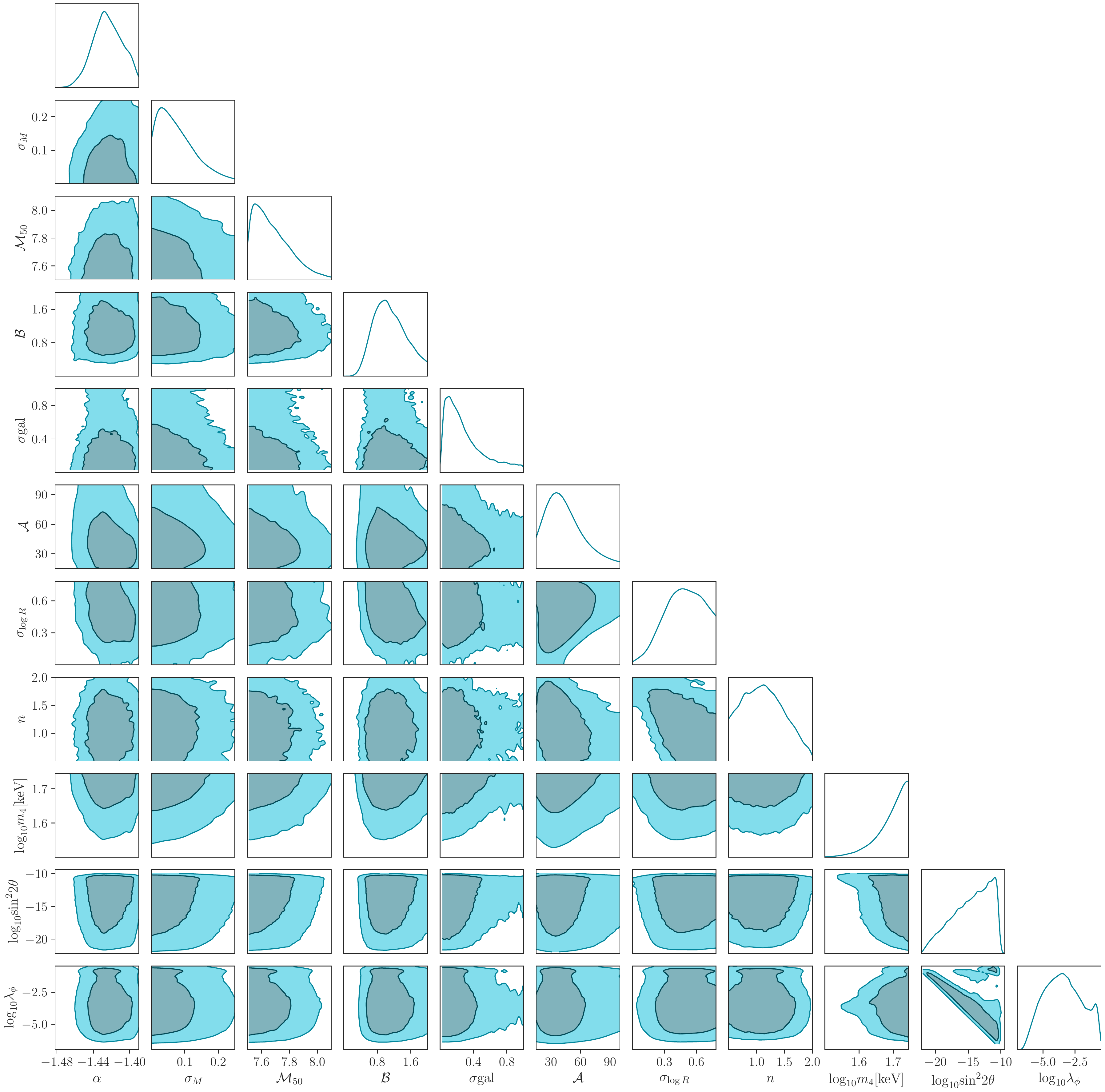}
\caption{Posterior distribution from the thermal-relic WDM fit to DES and PS1 MW satellites, presented in \cite{DES:2020fxi}, cast into the parameter space of our self-interacting sterile neutrino DM model (i.e., $m_4$, $\sin^2 2\theta$, and $\lambda_{\phi}$). Dark (light) shaded contours represent 68\% (95\%) confidence intervals, respectively. Marginalized posteriors are shown in the top panel of
each column. Note that $\sigma_M$, $\sigma_{\rm{gal}}$, and $\sigma_{\log R}$ are reported in $\rm{dex}$, $\mathcal{M}_{50}$ is reported as $\log_{10}(\mathcal{M}_{50}/\mathcal{M}_{\odot})$, $m_4$ is reported as $\log_{10}(m_4/\mathrm{keV})$, $\mathcal{A}$ is reported in $\rm{pc}$, and $\alpha$, $\mathcal{B}$, $n$, $\sin^2 2\theta$, and $\lambda_{\phi}$ are dimensionless. For details on the prior distributions assumed for galaxy--halo connection parameters when forward-modeling the MW satellite population, see \cite{DES:2020fxi}. }\label{fig:gh_snudm}
\end{figure*}

\appendix
Here, we present and discuss the posterior distribution resulting from our mapping of the self-interacting sterile neutrino DM model to a thermal-relic WDM model that was constrained by the MW satellite population in \cite{DES:2020fxi}. Figure~\ref{fig:gh_snudm} shows the posterior distribution over the eight galaxy--halo connection parameters, which are described in \cite{DES:2020fxi} and summarized in \cite{DES:2019ltu}, and our three self-interacting sterile neutrino DM parameters. The degeneracies between sterile neutrino DM parameters were discussed in the main text; here, we focus on degeneracies between sterile neutrino DM and the following galaxy--halo connection parameters:

\begin{enumerate}
    \item $\alpha$, the faint-end slope of the galaxy luminosity function. More negative values of $\alpha$ correspond to steeper luminosity functions, resulting in a larger number of predicted dwarf galaxies and thus a shallower stellar mass--halo mass relation in the context of abundance matching;
    \item $\sigma_M$, the scatter in satellite galaxy luminosity at fixed subhalo peak maximum circular velocity. Larger values of $\sigma_M$ cause more faint galaxies to up-scatter to observable luminosities and thus decreases the masses of the smallest halos inferred to host observed MW satellite galaxies;
    \item $\mathcal{M}_{50}$, the peak virial halo mass at which $50\%$ of halos host potentially observable satellite galaxies. In \cite{DES:2020fxi}, the galaxy occupation fraction is assumed to be a monotonic function of peak halo mass; thus, smaller values of $\mathcal{M}_{50}$ combined with small values of $\sigma_M$ ensure that the most massive subhalos host observable MW satellite galaxies (note that $\mathcal{M}_{50}$ and $\sigma_M$ are significantly anti-correlated in the Figure~\ref{fig:gh_snudm} posterior);
    \item $\mathcal{B}$, the efficiency of subhalo disruption due to the MW disk, defined relative to predictions from hydrodynamic simulations (\citealt{Garrison-Kimmel:2017zes}), such that $\mathcal{B}=1$ corresponds to hydrodynamic predictions and smaller (larger) $\mathcal{B}$ corresponds to less (more) efficient disruption (\citealt{Nadler:2017dxq,Nadler:2018iux}). Variations in $\mathcal{B}$ mostly affect the radial distribution and overall abundance of MW subhalos, rather than altering the subhalo population in a mass-dependent fashion;
    \item $\sigma_{\mathrm{gal}}$, the width of the galaxy occupation fraction. At fixed $\mathcal{M}_{50}$, increasing $\sigma_{\mathrm{gal}}$ results in a more gradual transition to galaxy-less halos, and thus allows lower-mass halos to contribute more significantly to the observed MW satellite population;
    \item $\mathcal{A}$, the amplitude of the relation between galaxy size (specifically, azimuthally averaged, projected half-light radius) and halo size (specifically, subhalo virial radius at the time of accretion onto the MW). For $\mathcal{A}\gtrsim 30~\mathrm{pc}$, increasing this parameter decreases the surface brightnesses of predicted satellites (at fixed luminosity), which makes bright satellites more difficult to detect---because they are generally closer to the surface brightness detectability boundary than dim satellites (\citealt{DES:2019vzn})---and forcing lower-mass halos to host observed systems. However, for sufficiently small values of $\mathcal{A}$, dim satellites are pushed below the size that \cite{DES:2019ltu,DES:2020fxi} assume to distinguish dwarf galaxies from star clusters (i.e., half-light radii of $10~\mathrm{pc}$), and thus forces more high-mass halos to contribute to the observed satellite population in the inference;
    \item $\sigma_{\log R}$, the scatter in satellite size at fixed halo size. Increasing $\sigma_{\log R}$ causes more small satellites to up-scatter to non-observable surface brightnesses, thus increasing the average mass of halos inferred to host MW satellites;
    \item $n$, the power-law index of the galaxy--halo size relation. Increasing $n$ yields a steeper size relation, such that low-mass halos host satellites decrease in size relative to high-mass halos. This shifts the halo population predicted to host observed MW satellites towards lower masses, because smaller satellites are generally easier to detect.
\end{enumerate}

Based on these effects, the degeneracies between sterile neutrino DM and galaxy--halo connection parameters are readily understood. In particular, any change to a galaxy--halo connection parameter that \emph{decreases} the average mass of halos inferred to host observed MW satellite galaxies results in a more stringent lower limit on $m_4$, because smaller values of $m_4$ preferentially suppress the abundance of low-mass halos, which are required to host observable galaxies in these regions of parameter space. Conversely, any change to a galaxy--halo connection parameter that \emph{increases} the average mass of halos inferred to host observed MW satellite galaxies allows for smaller values of $m_4$, because larger values of $m_4$ decrease the abundance of low-mass halos and thus do not manifest in observable changes to the predicted satellite population in these regions of parameter space. This reasoning explains all of the most prominent degeneracies between sterile neutrino DM and galaxy--halo connection parameters in Figure~\ref{fig:gh_snudm}; for example, $m_4$ is noticeably (and positively) correlated with $\sigma_M$, $\mathcal{M}_{50}$, and $\sigma_{\mathrm{gal}}$ because increasing any of these parameters decreases the average mass of halos that host observed MW satellites based on the parameter descriptions above.

Because the free-streaming scale is largely set by $m_4$ in our self-interacting sterile neutrino DM model, degeneracies between the remaining sterile neutrino DM and galaxy--halo connection parameters are much weaker. Note that $m_4$ is anti-correlated with both $\sin^2 2\theta$ and $\lambda_{\phi}$ in the posterior because self-interactions enable more frequent active neutrino scattering, allowing the DM relic density to be saturated at smaller mixing angles. Thus, these parameters all exhibit similar degeneracies with the galaxy--halo connection parameters.

\end{document}